A Review on Molecular Simulations for the Rupture of Polymer Networks

*Yuichi Masubuchi, Takato Ishida, Yusuke Koide, and Takashi Uneyama
Department of Materials Physics, Nagoya University, Nagoya 4649603, Japan

*To whom correspondence should be addressed: mas@mp.pse.nagoya-u.ac.jp

Ver Sep. 5, 2025

**Abstract**

Molecular simulations provide a powerful means to unravel the complex relationships between network architecture and the mechanical response of polymer networks, with a particular emphasis on rupture and fracture phenomena. Although simulation studies focused on polymer network rupture remain relatively limited compared to the broader field, recent advances have enabled increasingly nuanced investigations that bridge molecular structures and macroscopic failure behaviors. This review surveys the evolution of molecular simulation approaches for polymer network rupture, from early studies on related materials to state-of-the-art methods. Key challenges—including mismatched spatial and temporal scales with experiments, the validity of coarse-grained models, the choice of simulation protocols and boundary conditions, and the development of meaningful structural descriptors—are critically discussed. Special attention is paid to the assumptions underlying universality, limitations of current methodologies, and the ongoing need for theoretically sound and experimentally accessible network characterization. Continued progress in computational techniques, model development, and integration with experimental insights will be essential for a deeper, predictive understanding of polymer network rupture.

**1. Introduction**

The interplay between the molecular structure of polymer networks and their mechanical properties, particularly fracture and rupture, remains a fundamental yet elusive issue in polymer science [1,2]. To elucidate this complex relationship, molecular simulations have become increasingly indispensable tools [3]. As illustrated in Figure 1, the annual number of publications on polymer simulations (black curve) has grown exponentially in recent decades, exceeding 3,000 per year. Among these, studies focused on polymer networks (blue curve) follow a similar upward trajectory, while those explicitly investigating polymer network rupture and fracture (red curve)— though fewer—are steadily increasing.

Despite these advances, simulating polymer network rupture at the molecular level remains challenging due to the inherent complexity of network architectures and the scale disparities between simulations and experiments. This review begins with an overview of the historical development of rupture simulations, contextualizing polymer network studies within broader efforts on related materials. It then addresses the key conceptual and technical challenges faced in simulating rupture phenomena, highlighting recent progress, limitations, and open questions that define the current research frontier.

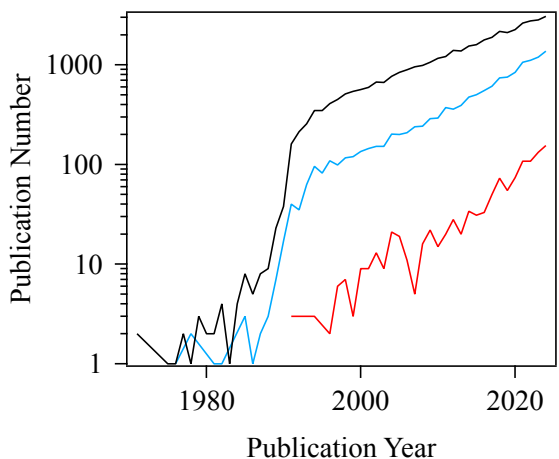

**Figure 1:** Annual number of publications on molecular simulations of polymers (black curve), polymer network simulations including rubbers, gels, and epoxies (blue curve), and simulations specifically addressing rupture and fracture of polymer networks (red curve), based on Web of Science data as of June 2025. The data illustrate exponential growth in polymer simulation research, with rupture-focused studies emerging as a distinct but still developing area.

## 2. Historical Background

As explained later, molecular-level simulations of polymer network rupture have predominantly emerged since the early 2000s. These efforts build upon foundational studies from related fields that provide critical insights into fracture phenomena.

Atomistic simulations of crack tips in crystalline solids date back to the 1970s [4]. Thomson et al.[5] revealed how atomic discreteness creates energy barriers for the propagation of cracks in brittle solids. Sinclair and Lawn [6,7] combined continuum elasticity with atomistic relaxation to model crack-tip structures in diamond-type crystals. Such a direction was followed by other researchers, who focused on atomic-scale mechanisms, such as crack-tip plasticity [8–10]. Note that some studies in the 1980s employed similar lattice setups as atomistic simulations, but for different aims. Herrmann et al. [11–13] and Duxbury et al. [14,15] investigated the random fuse model, whereas Meakin [16,17] studied spring network models to simulate crack nucleation and propagation in brittle, disordered solids, thereby laying the groundwork for network-based approaches to fracture. This direction, considering the fracture of the modeled elastic body, has been widely explored [18,19].

Parallel to fracture studies, polymer dynamics simulations began evolving in the 1970s [20–22]. To accommodate the slow dynamics, coarse-grained bead-spring models were employed from these earliest studies. In the late 1980s, Brownian dynamics simulation for the bead-spring chain was established to reproduce polymer dynamics in melts [23–26]. Molecular dynamics simulations with united atom models were also developed [27]. Building on these models and methodologies, studies have been conducted on the yield behaviors of polymeric glasses under elongation [28,29]. Later, owing to the progress in computational technologies, full-atomistic models have also been employed for glassy polymers. For instance, Hutnik et al. [30] reported full-atomistic simulations of polycarbonate under plastic deformations, based on the methodology established by Theodorou and Suter [31]. Recent computational facilities have enabled further large-scale and long-duration simulations [32,33].

In the 1990s, integrating the approaches mentioned above, Baijon and Robbins [34] introduced polymers into crack tip simulations to report apparent rupture of polymeric liquids. They placed melts of bead-spring chains between solid walls and observed the rupture of the melts as the distance between the walls increased, as shown in Fig. 2. Robbins et al. [35,36] extended this approach to the fracture of polymer glasses. Similar studies on polymer nanocomposites [37] and end-grafted polymers attached to the wall surface [38,39] have also been conducted.

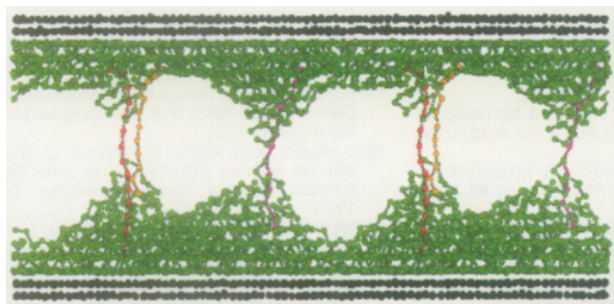

**Figure 2:** A snapshot of the melt rupture simulation between solid walls by Baijon and Robbins [34], with permission from the publisher.

Building on this foundation, Stevens [40,41] pioneered rupture simulations of densely cross-linked epoxy-like networks in the 2000s. He introduced bond breakage and varied the interfacial bonding density between polymers and solid walls, and observed the transition between cohesive and interfacial failure. (This cohesive failure corresponds to the rupture of the stretched polymer network between walls.) Following his work, attempts have been made to extend the model towards complex and realistic systems. Tsige et al. [42,43] assessed the influence of cross-linker functionality. Subsequent modifications introduced ionic interactions [44] and bending rigidity [45,46]. The effects of entanglement have also been discussed [47]. As a simulation study in the early period, the work by Yarovsky and Evans [48] is also noteworthy because they constructed a full-atomistic model of epoxy attached to an alumina surface and calculated the adhesion energy, although cohesive failure is not discussed.

Eliminating the effect of the wall boundary, Rottler and Robins [49,50] investigated the fracture of bead-spring polymers in the glassy state by applying boundary conditions that stretched the system. Following their method, Panico et al. [51] introduced crosslinks into glassy polymers to investigate the effects of crosslink density on fracture. With similar simulation settings, Nouri et al. [52] conducted full-atomistic simulations for the fracture of epoxy networks. Full-atomistic modeling was also attempted for polybutadiene rubber [53] and polyurethane [54]. Moller et al. [55] investigated epoxy employing a united atom model. For bead-spring models, the effects of bending rigidity [56], entanglements [57,58], chain stiffness [59], and loops [60] have been discussed. Large-scale bead-spring simulations have been reported for bimodal networks [61], polymer nanocomposites [62,63], double-network systems [64], and slide-ring networks [65]. Due to the widely dispersed relaxation modes, simulations for vitrimers have been attempted with further coarse-grained models [66,67].

Due to critical spatial and temporal scale challenges in molecular simulations, continuum approaches have been pursued concurrently. Early work by Tijssens et al.[68] modeled crazing in polymer glasses via a finite element method. Later, Miehe et al. [69,70] applied the phase field modeling technique to rubbery polymers, and this approach has been further explored [71–73]. Since this review focuses on molecular simulations, see recent reviews [74–76] on continuum approaches for further details.

Complementary to these continuum and atomistic approaches are mesoscopic models incorporating explicit polymer connectivity while simplifying other aspects. Arora et al. [77–79] introduced such a model to discuss the effects of topological defects, including loops and dangling ends, and spatial inhomogeneity of network node density. Masubuchi et al. [80–88] investigated similar phantom chain networks to discuss the effects of strand length, its bimodality, node functionality, conversion, prepolymer concentration, and other factors. A typical example is shown in Fig.3.

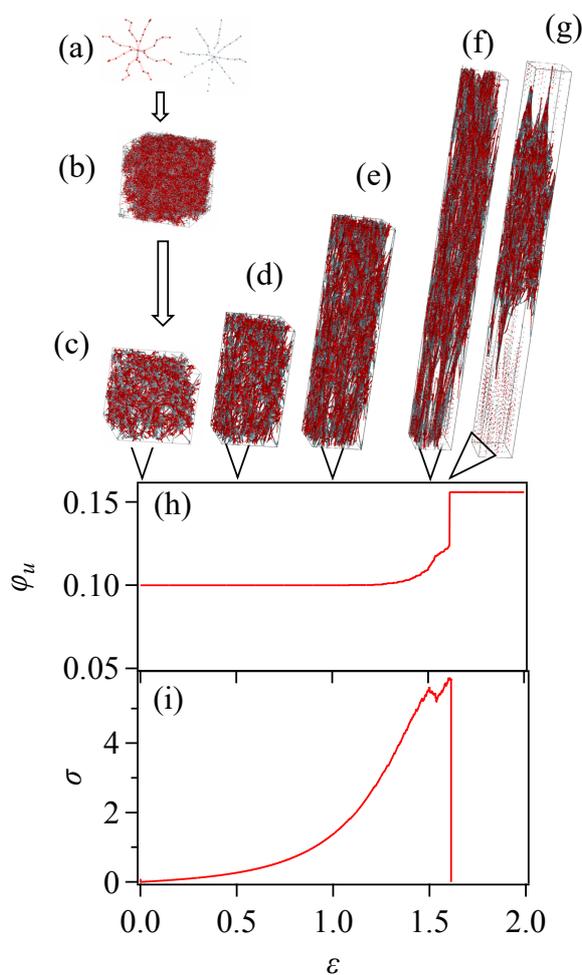

**Figure 3:** Typical snapshots in rupture simulations for phantom chain networks; prepolymers (a), the gelated network (b), the energy-minimized structure (c), the stretched states (d)-(f), the broken network (g), development of unconnected strand fraction $\varphi_u$ (h) and stress $\sigma$ (i) during the stretch plotted against true strain $\varepsilon$. The prepolymer functionality $f = 8$ and conversion $\varphi_c = 0.9$ taken from Masubuchi et al. [88], with permission from the publisher.

## 3. System size and stretching conditions

A critical aspect of molecular simulations of polymer network rupture is the choice of system size and simulation duration, which must be sufficiently large and long to capture the relevant phenomena. Due to computational constraints, the dimensions of simulations and strain rates often differ substantially from those of experimental conditions.

For example, Stevens [40,41] estimated the plastic zone size near a crack tip in epoxy to be approximately 10 μm, while his simulations with 170,000 beads represent a region smaller than 100 nm. More recent large-scale simulations with over 1.6 million beads [61] suggest that, at least for specific rupture characteristics, size effects may be limited; however, such generalizations depend heavily on the particular problem at hand. Notably, the simulation box size imposes artificial cutoffs on the probability distribution of fracture characteristics [89–92], a factor that is seldom discussed.

Temporal scaling presents an even greater challenge. In Stevens' work [40,41], the stretch speed used for most cases was $10^{-3}$ in Lennard-Jones (LJ) units, which is comparable to the Rouse relaxation rate of a linear chain with 30 beads [26,93], but significantly faster than the relaxation rate of his entire network containing 170,000 beads. The study by Sliozberg et al. [57] employed a stretching speed of $10^{-5}$ in LJ units for their system with 500,000 beads. Yet, they stated that this stretch is much faster than in experiments, as explicitly indicated in the title, "high-strain rate deformation", even for the systems including monomer beads as solvents. Even for recent simulations, the stretching speed remains to be higher than $10^{-5}$ in LJ units for most cases.

One may argue that the relaxation of the single strand is dominant in the relaxation of the network. This view is suitable for unbreakable rubbery networks [94–98]. In contrast, for network rupture and fracture, structural relaxation occurs after every single strand breakage. In a cascade of bond scission and macroscopic network failure, structural relaxation and mitigation propagate throughout the entire system, with a characteristic time that rapidly increases due to changes in network connectivity. For example, Brownian dynamics studies of phantom chain networks demonstrate that when strain rates exceed the reciprocal relaxation time of disconnected network domains, residual stresses persist even after macroscopic failure [84,85,87].

Note that most of the rupture simulations were made under constant stretch speed; the walls or the boundary of the simulation box are moved with a constant speed. This condition is consistent with most rupture experiments for polymeric solids and is fair when the effects of strain rate are negligible. In contrast, if the rupture behavior depends on the stretch speed reflecting the breakage of the network, deformation conditions under a constant Hencky strain rate would be appropriate for discussing the competition between relaxation and deformation, analogous to the extensional rheology of polymeric liquids. A few simulation studies explicitly state that they elongated the system with constant Hencky strain rates [60,84,85].

To alleviate the influence of strain rate, some studies employ quasi-static or energy minimization approaches that disregard dynamic effects, focusing instead on mechanical equilibrium and force balances [99–101]. Masubuchi et al. [80–83,86–88] employed this approach to observe network rupture, eliminating the effects of strain rate, as illustrated in Fig. 3. The drawback is the lack of relaxation and energy dissipation [60].

Another often unaddressed but essential factor is the choice of elongational boundary conditions [102]. In simulations with solid walls [40–47], simulation box sizes in the lateral directions are unchanged, and the volume increases as the system is stretched. As mentioned by Baijon and Robbins [34], these studies aim to reproduce what happens at the crack tip in tearing tests, where the system size increases as deformation is applied. Some simulations without solid walls also employ this condition [52,64,65]. The other approach is to determine the system size based on pressure using NPT ensemble techniques [51,53–56,60]. The remaining simulations assume incompressibility, and the simulation box sizes in the lateral directions decrease as the elongation increases [58,61,78,85,87]. These simulations aim to replicate the behavior of bulk materials under tensile testing conditions. Since tearing and tensile tests experimentally probe different failure mechanisms, careful consideration of boundary conditions is crucial for meaningful comparisons between simulations and experiments.

Lastly, the definition of stress employed in simulations affects the interpretation of stress–strain relations [102] . Experimental fracture testing commonly reports nominal (engineering) stress for convenience, whereas molecular simulations calculate true stress from microscopic virial expressions [103,104]. When lateral dimensions are fixed, nominal and true stresses coincide; otherwise, appropriate conversions are necessary to maintain correct conjugacy with nominal strain during data analysis [105–108].

## 4. Coarse-graining

As mentioned above, the coarse-grained bead-spring model and its derivatives have been utilized due to their efficiency in reducing computational costs. However, constructing and validating coarse-grained models is not a trivial task[109–113]. The widely adopted bead-spring model by Kremer and Grest [24] is justified by its ability to reproduce key features of entangled polymer dynamics, which exhibit universality across different chemistries as demonstrated by extensive experimental evidence [114,115]. This universality has enabled further coarse-grained descriptions, such as tube models, to effectively capture the dynamics of polymers [116–118].

Simulations of polymer networks often build on these insights, assuming that chemistry-dependent effects can be subsumed into a limited set of model parameters associated with beads and springs. However, the universality of rupture phenomena across diverse chemistries remains unestablished. Thus, the widespread rationalization of coarse-grained models for rupture remains pending. Fine-grained atomistic simulations [52,53,55] provide complementary insights, although they face even steeper challenges in bridging spatial and temporal scales.

A frequently underappreciated issue concerns the equation of motion under deformation within coarse-grained modeling. Projection operator techniques [119,120] demonstrate that the eliminated degrees of freedom in coarse-graining act as effective drag and random forces, leading to Langevin [104,121] or dissipative particle dynamics (DPD) [122,123] equations of motion. However, rigorous coarse-graining theory for nonequilibrium, deforming systems remains lacking. In addition, while several nonequilibrium molecular dynamics methods exist [124], no thermostat is yet theoretically proven to dissipate deformation-injected energy under strongly nonequilibrium conditions correctly.

Consequently, equations of motion used in rupture simulations vary. In particular, modeling the background flow in Langevin dynamics is inconsistent: some studies neglect it, assuming a quiescent flow, while others account for it [84,85]. Given that deformation rates in typical rupture simulations exceed reciprocal relaxation times, neglecting background flow may introduce artifacts, such as the suppression of inhomogeneous void formation. Complex flow patterns inevitably develop near voids and interfaces, further complicating the modeling. Modified DPD schemes [126,127] have been proposed to address these issues, with promising results demonstrated in liquid rupture simulations [125].

## 5. Network structure

A critical aspect of polymer network rupture simulations is the design and characterization of the network itself. Widely adopted approaches construct networks by mimicking experimental methods, such as the polymerization of small molecules [40–46,52,54–56,58,64], cross-linking of linear prepolymers in melts or solutions [47,51,53,57], end-linking of star polymer precursors [47,51,53,57], and cross-linking linear prepolymers with multifunctional linkers [59,78,82]. These methodologies largely build upon earlier foundational studies [128–130].

However, the network structures generated in simulations may differ from actual experimental materials due to inherent challenges in replicating reaction kinetics and gelation timescales. Kinetic arrest during gelation [131,132] commonly alters the network topology, and simulating these dynamic processes accurately is challenging due to the mismatched time domains between simulations and experiments. Consequently, rigorous evaluation of the simulated network structure is necessary. This task is inherently circular: dominant structural descriptors for rupture should guide network design, yet identifying such descriptors often requires analyzing the network post-simulation. Moreover, experimental characterization of network topology remains an evolving field [133].

Assuming the created networks approximate experimental systems, efforts have focused on identifying the key structural parameters that govern rupture. Early work by Stevens [40,41] emphasized the shortest path in the stretching direction as an important descriptor, a concept recently refined by Yu and Jackson [134]. Rooted in the Lake-Thomas theory [135], classical parameters such as network node density and node functionality continue to play central roles in network analysis. Extending this approach, Barney et al. [60] investigated the influence of loop fractions on fracture energy, connecting simulations to experiments via the theoretical model [136]. Recently, cycle rank—a topological quantity representing the density of independent loops—has been proposed as an effective descriptor that unifies influences of node functionality and conversion [88]. This metric is amenable to estimation using mean-field theories [137,138], enabling experimental applicability [139]. Yang and Qu [56] discussed the formation of cavities in epoxy in relation to rupture. Zhang and Riggleman [140] investigated network failure using geodesic edge betweenness centrality, building on previous studies in 2D systems [141,142].

Most of these studies implicitly assume a degree of universality across chemistries by employing coarse-grained models. For instance, the rupture characteristics for phantom chain networks with varying node functionalities and conversions, but identical strand lengths, collapse onto master curves when plotted against cycle rank density [82,88] (Fig. 4). Although promising, this universality remains unverified for chemically diverse systems.

Besides, there is an open problem regarding universality across structurally distinct network classes. For example, networks based on regular lattice topologies or graph theory may exhibit rupture behavior that differs fundamentally from that of random or statistically generated networks, highlighting the need for further investigation [143–145].

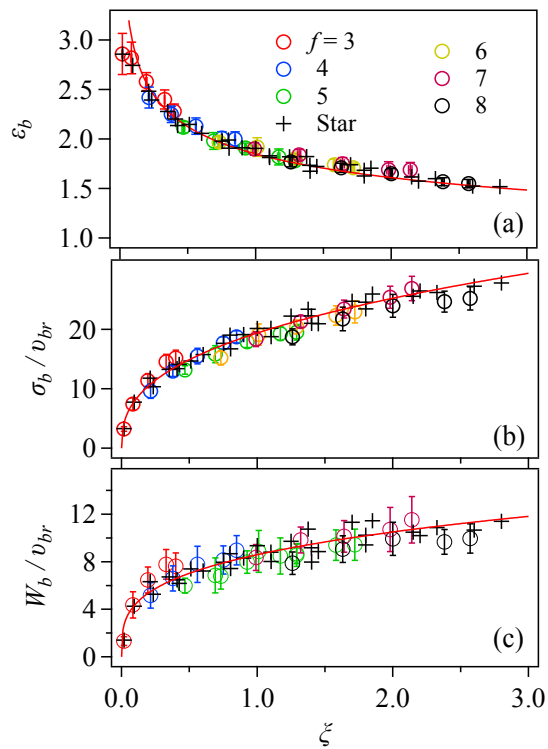

**Figure 4** Strain at break $\varepsilon_b$ (a), stress at break $\sigma_b$ (b), and work for rupture $W_b$ (c) plotted against cycle rank $\xi$ for phantom chain networks with various node functionality $f$ and conversions between 0.6 and 0.95. The strand segment number is 10, and the strand density is 8. The circles show the results from the systems created from linear prepolymers and multi-functional linkers, whereas the cross shows those for star prepolymers.

## 6. Summary

Molecular simulations have become indispensable for unraveling the rupture behavior of polymer networks, offering microscopic insight into phenomena that are challenging to access experimentally. Nevertheless, progress in this field continues to be hampered by several fundamental difficulties. One persistent challenge is the significant mismatch in spatial and temporal scales between simulations and experiments. While advances in computational power and coarse-grained modeling have enabled larger and longer simulations, key assumptions—such as the universality of rupture behavior across different chemistries—have yet to be systematically validated. This limitation is particularly significant for rupture phenomena, as local chemistry and network topology can both profoundly influence fracture response. Equally important is the careful selection and transparent reporting of simulation parameters, including system size, deformation protocol, boundary conditions, and the definitions of stress and strain. These factors critically impact the interpretation of simulation results and their applicability to real-world applications. Similarly, the impact of coarse-graining strategies and choices of equations of motion or thermostats in nonequilibrium conditions must be scrutinized for their effect on the fidelity of rupture simulations. A major unresolved issue is the development and validation of robust, theoretically sound, and experimentally accessible descriptors of network structure. Without such descriptors, direct and meaningful comparison between simulations and experimental systems remains elusive. In sum, while molecular simulations have shed light on the rupture of polymer networks, continued progress will require both methodological innovations and more precise, theory-driven characterization techniques to close the gap between model predictions and experimental observations.

**Acknowledgements**

The Hibi Foundation partly supported this study.

REFERENCES

[1]     Gu Y, Zhao J and Johnson J A 2020 Polymer Networks:From Plastics and Gels to Porous Frameworks *Angewandte Chemie* **132** 5054–85

[2]     Danielsen S P O, Beech H K, Wang S, El-Zaatari B M, Wang X, Sapir L, Ouchi T, Wang Z, Johnson P N, Hu Y, Lundberg D J, Stoychev G, Craig S L, Johnson J A, Kalow J A, Olsen B D and Rubinstein M 2021 Molecular Characterization of Polymer Networks *Chem Rev* **121** 5042–92

[3]     Tauber J, van der Gucht J and Dussi S 2022 Stretchy and disordered: Toward understanding fracture in soft network materials via mesoscopic computer simulations *J Chem Phys* **156** 160901

[4]     Patil S P and Heider Y 2019 A Review on Brittle Fracture Nanomechanics by All-Atom Simulations *Nanomaterials* **9** 1050

[5]     Thomson R, Hsieh C and Rana V 1971 Lattice Trapping of Fracture Cracks *J Appl Phys* **42** 3154–60

[6]     Sinclair J E 1972 Atomistic computer simulation of brittle-fracture extension and closure *Journal of Physics C: Solid State Physics* **5** L271–4

[7]     Sinclair J E and Lawn B R 1972 An atomistic study of cracks in diamond-structure crystals *Proceedings of the Royal Society of London. A. Mathematical and Physical Sciences* **329** 83–103

[8]     Ashurst W T and Hoover W G 1976 Microscopic fracture studies in the two-dimensional triangular lattice *Phys Rev B* **14** 1465–73

[9]     Andric P and Curtin W A 2019 Atomistic modeling of fracture *Model Simul Mat Sci Eng* **27** 013001

[10]   Lawn B R and Marshall D B 2022 Brittle Solids: From Physics and Chemistry to Materials Applications *Annu Rev Mater Res* **52** 441–71

[11]   de Arcangelis L, Redner S and Herrmann H J 1985 A random fuse model for breaking processes *Journal de Physique Lettres* **46** 585–90

[12]   Hansen A, Roux S and Herrmann H J 1989 Rupture of central-force lattices *Journal de Physique* **50** 733–44

[13]   Herrmann H J, Hansen A and Roux S 1989 Fracture of disordered, elastic lattices in two dimensions *Phys Rev B* **39** 637–48

[14]   Duxbury P M, Kim S G and Leath P L 1994 Size effect and statistics of fracture in random materials *Materials Science and Engineering: A* **176** 25–31

[15]   Duxbury P M, Leath P L and Beale P D 1987 Breakdown properties of quenched random systems: The random-fuse network *Phys Rev B* **36** 367–80

[16]   Meakin P 1987 A simple model for elastic fracture in thin films *Thin Solid Films* **151** 165–90

[17]   Meakin P, Li G, Sander L M, Louis E and Guinea F 1989 A simple two-dimensional model for crack propagation *J Phys A Math Gen* **22** 1393–403

[18]   Patinet S, Vandembroucq D, Hansen A and Roux S 2014 Cracks in random brittle solids: *Eur Phys J Spec Top* **223** 2339–51

[19]   Wiese K J 2022 Theory and experiments for disordered elastic manifolds, depinning, avalanches, and sandpiles *Reports on Progress in Physics* **85** 086502

[20]   Rapaport D C 1978 *Molecular dynamics simulation of polymer chains with excluded volume* vol 11

[21]   Fixman M 1978 Simulation of polymer dynamics. II. Relaxation rates and dynamic viscosity *J Chem Phys* **69** 1538

[22]   Fixman M 1978 Simulation of polymer dynamics. I. General theory *J Chem Phys* **69** 1527

[23]   Everaers R, Karimi-Varzaneh H A, Fleck F, Hojdis N and Svaneborg C 2020 Kremer–Grest Models for Commodity Polymer Melts: Linking Theory, Experiment, and Simulation at the Kuhn Scale *Macromolecules* **53** 1901–16

[24]   Kremer K and Grest G S 1990 Dynamics of entangled linear polymer melts: A molecular-dynamics simulation *J Chem Phys* **92** 5057

[25]   Grest G S and Kremer K 1986 Molecular dynamics simulation for polymers in the presence of a heat bath *Phys Rev A   (Coll Park)* **33** 3628–31

[26]   Kremer K, Grest G S and Carmesin I 1988 Crossover from rouse to reptation dynamics: A molecular-dynamics simulation *Phys Rev Lett* **61** 566–9

[27]   Rigby D and Roe R J 1987 Molecular dynamics simulation of polymer liquid and

glass. I. Glass transition *J Chem Phys* **87** 7285–92

[28]     Brown D and Clarke J H R 1991 Molecular dynamics simulation of an amorphous polymer under tension. 1. Phenomenology *Macromolecules* **24** 2075–82

[29]     Rottler J and Robbins M O 2001 Yield conditions for deformation of amorphous polymer glasses *Phys Rev E* **64** 051801

[30]     Hutnik M, Argon A S and Suter U W 1993 Simulation of elastic and plastic response in the glassy polycarbonate of 4,4'-isopropylidenediphenol *Macromolecules* **26** 1097–108

[31]     Theodorou D N and Suter U W 1986 Atomistic Modeling of Mechanical Properties of Polymeric Glasses *Macromolecules* **19** 139–54

[32]     Fujimoto K, Tang Z, Shinoda W and Okazaki S 2019 All-atom molecular dynamics study of impact fracture of glassy polymers. I: Molecular mechanism of brittleness of PMMA and ductility of PC *Polymer (Guildf)* **178** 121570

[33]     Fujimoto K 2022 Fracture and Toughening Mechanisms of Glassy Polymer at the Molecular Level *Nihon Reoroji Gakkaishi* **50** 37–41

[34]     Baljon A R C and Robbins M O 1996 Energy Dissipation During Rupture of Adhesive Bonds *Science (1979)* **271** 482–4

[35]     Gersappe D and Robbins M O 1999 Where do polymer adhesives fail? *Europhysics Letters (EPL)* **48** 150–5

[36]     Baljon A R C and Robbins M O 2001 Simulations of Crazing in Polymer Glasses:   Effect of Chain Length and Surface Tension *Macromolecules* **34** 4200–9

[37]     Gersappe D 2002 Molecular Mechanisms of Failure in Polymer Nanocomposites *Phys Rev Lett* **89** 058301

[38]     Sides S W, Grest G S, Stevens M J and Plimpton S J 2004 Effect of end‐tethered polymers on surface adhesion of glassy polymers *J Polym Sci B Polym Phys* **42** 199–208

[39]     Morita H, Yamada M, Yamaguchi T and Doi M 2005 Molecular Dynamics Study of the Adhesion between End-grafted Polymer Films *Polym J* **37** 782–8

[40]     Stevens M J 2001 Manipulating connectivity to control fracture in network polymer adhesives *Macromolecules* **34** 1411–5

[41]     Stevens M J 2001 Interfacial Fracture between Highly Cross-Linked Polymer Networks and a Solid Surface:   Effect of Interfacial Bond Density *Macromolecules* **34** 2710–8

[42]     Tsige M and Stevens M J 2004 Effect of cross-linker functionality on the adhesion of highly cross-linked polymer networks: A molecular dynamics study of epoxies *Macromolecules* **37** 630–7

[43]   Tsige M, Lorenz C D and Stevens M J 2004 Role of network connectivity on the mechanical properties of highly cross-linked polymers *Macromolecules* **37** 8466–72

[44]   Dirama T E, Varshney V, Anderson K L, Shumaker J A and Johnson J A 2008 Coarse-grained molecular dynamics simulations of ionic polymer networks *Mech Time Depend Mater* **12** 205–20

[45]   Mukherji D and Abrams C F 2008 Microvoid formation and strain hardening in highly cross-linked polymer networks *Phys Rev E Stat Nonlin Soft Matter Phys* **78**

[46]   Mukherji D and Abrams C F 2009 Mechanical behavior of highly cross-linked polymer networks and its links to microscopic structure *Phys Rev E Stat Nonlin Soft Matter Phys* **79**

[47]   Solar M, Qin Z and Buehler M J 2014 Molecular mechanics and performance of crosslinked amorphous polymer adhesives *J Mater Res* **29** 1077–85

[48]   Yarovsky I 2002 Computer simulation of structure and properties of crosslinked polymers: application to epoxy resins *Polymer (Guildf)* **43** 963–9

[49]   Rottler J, Barsky S and Robbins M O 2002 Cracks and Crazes: On Calculating the Macroscopic Fracture Energy of Glassy Polymers from Molecular Simulations *Phys Rev Lett* **89**

[50]   Rottler J and Robbins M O 2003 Growth, microstructure, and failure of crazes in glassy polymers *Phys Rev E* **68** 011801

[51]   Panico M, Narayanan S and Brinson L C 2010 Simulations of tensile failure in glassy polymers: effect of cross-link density *Model Simul Mat Sci Eng* **18** 055005

[52]   Nouri N and Ziaei-Rad S 2011 A molecular dynamics investigation on mechanical properties of cross-linked polymer networks *Macromolecules* **44** 5481–9

[53]   Payal R S, Fujimoto K, Jang C, Shinoda W, Takei Y, Shima H, Tsunoda K and Okazaki S 2019 Molecular mechanism of material deformation and failure in butadiene rubber: Insight from all-atom molecular dynamics simulation using a bond breaking potential model *Polymer (Guildf)* **170** 113–9

[54]   Zhang C, Li Y, Wu Y, Wang C, Liang J, Xu Z, Zhao P and Wang J 2024 Key Role of Cross-Linking Homogeneity in Polyurethane Mechanical Properties: Insights from Molecular Dynamics *J Phys Chem B* **128** 12612–27

[55]   Moller J C, Barr S A, Schultz E J, Breitzman T D and Berry R J 2013 Simulation of Fracture Nucleation in Cross-Linked Polymer Networks *JOM* **65** 147–67

[56]   Yang S and Qu J 2014 Coarse-grained molecular dynamics simulations of the tensile behavior of a thermosetting polymer *Phys Rev E Stat Nonlin Soft Matter Phys* **90**

[57]   Sliozberg Y R, Hoy R S, Mrozek R A, Lenhart J L and Andzelm J W 2014 Role of entanglements and bond scission in high strain-rate deformation of polymer gels

*Polymer (Guildf)* **55** 2543–51

[58]  Furuya T and Koga T 2025 Molecular Simulation of Effects of Network Structure on Fracture Behavior of Gels Synthesized by Radical Polymerization *Macromolecules* **58** 3359–68

[59]  Zheng X, Xia W and Zhang Y 2024 Understanding the role of chain stiffness in the mechanical response of cross-linked polymer: Flexible vs. semi-flexible chains *Extreme Mech Lett* **73**

[60]  Barney C W, Ye Z, Sacligil I, McLeod K R, Zhang H, Tew G N, Riggleman R A and Crosby A J 2022 Fracture of model end-linked networks *Proceedings of the National Academy of Sciences* **119** 2–7

[61]  Hagita K and Murashima T 2024 Critical Importance of Both Bond Breakage and Network Heterogeneity in Hysteresis Loop on Stress–Strain Curves and Scattering Patterns *Macromolecules* **57** 10903–11

[62]  Hagita K, Morita H and Takano H 2016 Molecular dynamics simulation study of a fracture of filler-filled polymer nanocomposites *Polymer (Guildf)* **99** 368–75

[63]  David A, Tartaglino U, Casalegno M and Raos G 2021 Fracture in Silica/Butadiene Rubber: A Molecular Dynamics View of Design–Property Relationships *ACS Polymers Au* **1** 175–86

[64]  Higuchi Y, Saito K, Sakai T, Gong J P and Kubo M 2018 Fracture Process of Double-Network Gels by Coarse-Grained Molecular Dynamics Simulation *Macromolecules* **51** 3075–87

[65]  Uehara S, Wang Y, Ootani Y, Ozawa N and Kubo M 2022 Molecular-Level Elucidation of a Fracture Process in Slide-Ring Gels via Coarse-Grained Molecular Dynamics Simulations *Macromolecules* **55** 1946–56

[66]  Raffaelli C, Bose A, Vrusch C H M P, Ciarella S, Davris T, Tito N B, Lyulin A V., Ellenbroek W G and Storm C 2020 Rheology, Rupture, Reinforcement and Reversibility: Computational Approaches for Dynamic Network Materials *Self-Healing and Self-Recovering Hydrogels* ed C Creton and O Okay (Cham: Springer International Publishing) pp 63–126

[67]  Tito N B, Creton C, Storm C and Ellenbroek W G 2019 Harnessing entropy to enhance toughness in reversibly crosslinked polymer networks *Soft Matter* **15** 2190–203

[68]  Tijssens M G A, Giessen E van der and Sluys L J 2000 Simulation of mode I crack growth in polymers by crazing *Int J Solids Struct* **37** 7307–27

[69]  Miehe C and Schänzel L-M 2014 Phase field modeling of fracture in rubbery polymers. Part I: Finite elasticity coupled with brittle failure *J Mech Phys Solids* **65**

93–113

[70]   Miehe C, Hofacker M, Schänzel L-M and Aldakheel F 2015 Phase field modeling of fracture in multi-physics problems. Part II. Coupled brittle-to-ductile failure criteria and crack propagation in thermo-elastic–plastic solids *Comput Methods Appl Mech Eng* **294** 486–522

[71]   Li B and Bouklas N 2020 A variational phase-field model for brittle fracture in polydisperse elastomer networks *Int J Solids Struct* **182–183** 193–204

[72]   Yin B and Kaliske M 2020 Fracture simulation of viscoelastic polymers by the phase-field method *Comput Mech* **65** 293–309

[73]   Arunachala P K, Abrari Vajari S, Neuner M and Linder C 2023 A multiscale phase field fracture approach based on the non-affine microsphere model for rubber-like materials *Comput Methods Appl Mech Eng* **410** 115982

[74]   Zhang M, Zainal Abidin A R and Tan C S 2024 State-of-the-art review on meshless methods in the application of crack problems *Theoretical and Applied Fracture Mechanics* **131** 104348

[75]   Zhang G, Tang C, Chen P, Long G, Cao J and Tang S 2023 Advancements in Phase-Field Modeling for Fracture in Nonlinear Elastic Solids under Finite Deformations *Mathematics* **11** 3366

[76]   Alshoaibi A M and Fageehi Y A 2024 Advances in Finite Element Modeling of Fatigue Crack Propagation *Applied Sciences* **14** 9297

[77]   Arora A 2025 Effect of Spatial Heterogeneity on the Elasticity and Fracture of Polymer Networks *Macromolecules* **58** 1143–55

[78]   Arora A, Lin T S and Olsen B D 2022 Coarse-Grained Simulations for Fracture of Polymer Networks: Stress Versus Topological Inhomogeneities *Macromolecules* **55** 4–14

[79]   Arora A, Lin T-S, Beech H K, Mochigase H, Wang R and Olsen B D 2020 Fracture of Polymer Networks Containing Topological Defects *Macromolecules* **53** 7346–55

[80]   Masubuchi Y 2024 Phantom chain simulations for fracture of polymer networks created from star polymer mixtures of different functionalities *Polym J* **56** 163–71

[81]   Masubuchi Y 2025 Phantom Chain Simulations for the Fracture of Star Polymer Networks on the Effect of Arm Molecular Weight *Macromolecules* **58** 6399–405

[82]   Masubuchi Y 2024 Phantom chain simulations for fracture of end-linking networks *Polymer (Guildf)* **297** 126880

[83]   Masubuchi Y 2024 Phantom Chain Simulations for the Effect of Stoichiometry on the Fracture of Star-Polymer Networks *Nihon Reoroji Gakkaishi* **52** 21–6

[84]   Masubuchi Y, Koide Y, Ishida T and Uneyama T 2025 Brownian simulations for

fracture of star polymer phantom networks *Polym J* **57** 483–9

[85]   Masubuchi Y, Yamazaki R, Doi Y, Uneyama T, Sakumichi N and Sakai T 2022 Brownian simulations for tetra-gel-type phantom networks composed of prepolymers with bidisperse arm length *Soft Matter* **18** 4715–24

[86]   Masubuchi Y, Ishida T, Koide Y and Uneyama T 2024 Phantom chain simulations for the fracture of star polymer networks with various strand densities *Soft Matter* **20** 7103–10

[87]   Masubuchi Y, Doi Y, Ishida T, Sakumichi N, Sakai T, Mayumi K and Uneyama T 2023 Phantom Chain Simulations for the Fracture of Energy-Minimized Tetra- and Tri-Branched Networks *Macromolecules* **56** 2217–23

[88]   Masubuchi Y, Doi Y, Ishida T, Sakumichi N, Sakai T, Mayumi K, Satoh K and Uneyama T 2023 Phantom-Chain Simulations for the Effect of Node Functionality on the Fracture of Star-Polymer Networks *Macromolecules* **56** 9359–67

[89]   Daniels H E 1945 The statistical theory of the strength of bundles of threads. I *Proc R Soc Lond A Math Phys Sci* **183** 405–35

[90]   Bažant Z P 2004 Probability distribution of energetic-statistical size effect in quasibrittle fracture *Probabilistic Engineering Mechanics* **19** 307–19

[91]   Weibull W 1951 A statistical distribution function of wide applicability *J Appl Mech* **18** 293–7

[92]   Okabe T and Takeda N 2002 Size effect on tensile strength of unidirectional CFRP composites— experiment and simulation *Compos Sci Technol* **62** 2053–64

[93]   Masubuchi Y, Takata H, Amamoto Y and Yamamoto T 2018 Relaxation of Rouse Modes for Unentangled Polymers Obtained by Molecular Simulations *Nihon Reoroji Gakkaishi* **46** 171–8

[94]   Everaers R, Kremer K and Grest G S 1995 Entanglement effects in model polymer networks *Macromol Symp* **93** 53–67

[95]   Grest G S, Pütz M, Everaers R and Kremer K 2000 Stress–strain relation of entangled polymer networks *J Non Cryst Solids* **274** 139–46

[96]   Duering E R, Kremer K and Grest G S 1994 Structure and relaxation of end-linked polymer networks *J Chem Phys* **101** 8169

[97]   Duering E R, Kremer K and Grest G S 1993 Dynamics of model networks: the role of the melt entanglement length *Macromolecules* **26** 3241–4

[98]   Grest G S, Kremer K and Duering E R 1993 Kinetics and relaxation of end crosslinked polymer networks *Physica A: Statistical Mechanics and its Applications* **194** 330–7

[99]   Nishi K, Noguchi H, Sakai T and Shibayama M 2015 Rubber elasticity for

percolation network consisting of Gaussian chains *J Chem Phys* **143** 184905

[100] Nishi K, Chijiishi M, Katsumoto Y, Nakao T, Fujii K, Chung U, Noguchi H, Sakai T and Shibayama M 2012 Rubber elasticity for incomplete polymer networks *J Chem Phys* **137** 224903

[101] Gusev A A 2019 Numerical Estimates of the Topological Effects in the Elasticity of Gaussian Polymer Networks and Their Exact Theoretical Description *Macromolecules* **52** 3244–51

[102] Masubuchi Y, Ishida T, Koide Y and Uneyama T 2025 Influence of Stretching Boundary Conditions on Fracture in Phantom Star Polymer Networks: From Volume to Cross-sectional Area Conservation

[103] Irving J H and Kirkwood J G 1950 The Statistical Mechanical Theory of Transport Processes. IV. The Equations of Hydrodynamics *J Chem Phys* **18** 817–29

[104] Allen M P and Tildesley D J 1987 *Computer Simulation of Liquids* (Oxford: Clarendon Press)

[105] Henchy H 1928 Uber die Form des Elastizitatsgesetzes bei ideal elastischen Stoffen *Zeit. Tech. Phys.* **9** 215–20

[106] Curnier A and Rakotomanana L 1991 Generalized Strain and Stress Measures : Critical Survey and New Results *Engineering Transactions* **39** 461–538

[107] Korobeinikov S N 2001 Natural stress tensors *Journal of Applied Mechanics and Technical Physics* **42** 1051–6

[108] Norris A 2008 Eulerian conjugate stress and strain *J Mech Mater Struct* **3** 243–60

[109] Müller-Plathe F 2002 Coarse-Graining in Polymer Simulation: From the Atomistic to the Mesoscopic Scale and Back *ChemPhysChem* **3** 754–69

[110] de Pablo J J 2011 Coarse-grained simulations of macromolecules: from DNA to nanocomposites. *Annu. Rev. Phys. Chem.* **62** 555–74

[111] Kinjo T and Hyodo S 2007 Equation of motion for coarse-grained simulation based on microscopic description *Phys Rev E* **75** 051109

[112] Shi R, Qian H and Lu Z 2023 Coarse‐grained molecular dynamics simulation of polymers: Structures and dynamics *WIREs Computational Molecular Science* **13**

[113] Padding J T and Briels W J 2007 Ab-initio Coarse-Graining of Entangled Polymer Systems *Nanostructured Soft Matter* pp 437–60

[114] Ferry J D 1980 *Viscoelastic Properties of Polymers* (John Wiley & Sons, Inc.)

[115] Rubinstein M and Colby R H 2003 *Polymer Physics* (Oxford University PressOxford)

[116] Doi M and Edwards S F 1986 *The Theory of Polymer Dynamics* (Clarendon: Oxford University Press)

[117] Masubuchi Y 2014 Simulating the Flow of Entangled Polymers *Annu Rev Chem Biomol Eng* **5** 11–33

[118] Masubuchi Y 2016 Molecular Modeling for Polymer Rheology *Reference Module in Materials Science and Materials Engineering* (Elsevier) pp 1–7

[119] Kawasaki K 1973 Simple derivations of generalized linear and nonlinear Langevin equations *Journal of Physics A: Mathematical, Nuclear and General* **6** 1289–95

[120] Grabert H 1982 The projection operator technique pp 9–26

[121] Öttinger H C 2007 Systematic Coarse Graining: "Four Lessons and A Caveat" from Nonequilibrium Statistical Mechanics *MRS Bull* **32** 936–40

[122] Groot R D and Warren P B 1997 Dissipative particle dynamics: Bridging the gap between atomistic and mesoscopic simulation *Journal Of Chemical Physics* **107** 4423–35

[123] Hoogerbrugge P J and Koelman J M V A 1992 Simulating Microscopic Hydrodynamic Phenomena with Dissipative Particle Dynamics *Europhysics Letters (EPL)* **19** 155–60

[124] Evans D J and Morriss G 2008 *Statistical Mechanics of Nonequilibrium Liquids* (Cambridge University Press)

[125] Warren P B 2003 Vapor-liquid coexistence in many-body dissipative particle dynamics *Phys Rev E* **68** 066702

[126] Español P and Warren P 1995 Statistical Mechanics of Dissipative Particle Dynamics *Europhysics Letters (EPL)* **30** 191–6

[127] Zhao J, Chen S, Zhang K and Liu Y 2021 A review of many-body dissipative particle dynamics (MDPD): Theoretical models and its applications *Physics of Fluids* **33**

[128] Li C and Strachan A 2010 Molecular simulations of crosslinking process of thermosetting polymers *Polymer (Guildf)* **51** 6058–70

[129] Mavrantzas V G 2021 Using Monte Carlo to Simulate Complex Polymer Systems: Recent Progress and Outlook *Front Phys* **9**

[130] Yamamoto S and Tanaka K 2025 Molecular Dynamics Simulation of Cross‐linked Epoxy Resins: Past and Future *Macromol Rapid Commun* **46**

[131] Sirk T W 2020 Growth and arrest of topological cycles in small physical networks *Proceedings of the National Academy of Sciences* **117** 15394–6

[132] Jang C, Sirk T W, Andzelm J W and Abrams C F 2015 Comparison of Crosslinking Algorithms in Molecular Dynamics Simulation of Thermosetting Polymers *Macromol Theory Simul* **24** 260–70

[133] Gu Y, Zhao J and Johnson J A 2019 A (Macro)Molecular-Level Understanding of Polymer Network Topology *Trends Chem* **1** 318–34

[134] Yu Z and Jackson N E 2025 Shortest Paths Govern Bond Rupture in Thermoset Networks *Macromolecules* **58** 1728–36

[135] Lake J and Thomas A G 1967 The strength of highly elastic materials *Proc R Soc Lond A Math Phys Sci* **300** 108–19

[136] Zhong M, Wang R, Kawamoto K, Olsen B D and Johnson J A 2016 Quantifying the impact of molecular defects on polymer network elasticity *Science (1979)* **353** 1264–8

[137] Miller D R and Macosko C W 1976 A New Derivation of Post Gel Properties of Network Polymers *Macromolecules* **9** 206–11

[138] Macosko C W and Miller D R 1976 A New Derivation of Average Molecular Weights of Nonlinear Polymers *Macromolecules* **9** 199–206

[139] Matsuda T, Kashimura N, Sakai T, Nakajima T, Matsushita T and Gong J P 2025 Yielding of Double-Network Hydrogels with Systematically Controlled Tetra-PEG First Networks *Macromolecules* **58** 6017–32

[140] Zhang H and Riggleman R A 2024 Predicting failure locations in model end-linked polymer networks *Phys Rev Mater* **8** 035604

[141] Berthier E, Porter M A and Daniels K E 2019 Forecasting failure locations in 2-dimensional disordered lattices *Proceedings of the National Academy of Sciences* **116** 16742–9

[142] Pournajar M, Zaiser M and Moretti P 2022 Edge betweenness centrality as a failure predictor in network models of structurally disordered materials *Sci Rep* **12** 11814

[143] Lang M, Michalke W and Kreitmeier S 2001 A statistical model for the length distribution of meshes in a polymer network *J Chem Phys* **114** 7627–32

[144] Michalke W, Lang M, Kreitmeier S and Göritz D 2002 Comparison of topological properties between end-linked and statistically cross-linked polymer networks *J Chem Phys* **117** 6300–7

[145] Steger A and Wormald N C 1999 Generating Random Regular Graphs Quickly *Combinatorics, Probability and Computing* **8** 377–96